\documentclass{article}
\usepackage{graphicx} 
\usepackage{siunitx}
\usepackage{setspace}
\usepackage{array}
\usepackage{authblk}
\usepackage{xcolor}
\usepackage[style = authoryear, maxcitenames = 1]{biblatex} 
\usepackage{multirow}
\usepackage{geometry} 
\usepackage{booktabs}
\usepackage{float}
\usepackage{ulem} 

\title{Bunting and the ghost runner: a causal inference approach}

\author{Kevin Cummiskey}
\author{Lucas Villanti}
\author{Ira Crofford}

\affil{Department of Mathematical Sciences\\United States Military Academy, West Point, NY}

\begin{document}

\maketitle

\begin{abstract}
In this paper, we investigate the effectiveness of the home team bunting in extra innings of Major League Baseball games when the game is tied in the bottom of the inning. Using methods rooted in causal inference, we show that teams choose not to bunt when the statistical evidence clearly suggests it is in their advantage to do so. The reluctance to bunt in this situation is likely tied to the general decline in bunting, which is considered less effective than swinging away in most situations. In the 2021-22 seasons, the home team's first batter bunted in only 21\% of tied extra innings.  When they bunted, the home team went on to win the game in 74\% of those innings, compared to 57\% when they swung away (i.e. did not bunt), for an odds ratio for winning comparing bunting to swinging away of 2.13 (95\% C.I.: 1.13, 4.30).  However, the bunters had more experience bunting and the nonbunters were stronger hitters, thus the odds ratio above is a confounded estimate of the effect of bunting. Using inverse probability weighting to adjust for confounding, we estimate an odds ratio of 1.86 (95\% C.I. 1.07, 3.27).  This means that a typical team would expect to win 2 more games each season if they bunted in such situations which is worth millions in player salary.  
\end{abstract}

\newpage

\section{Introduction}

Beginning in 2020, Major League Baseball (MLB) implemented the most significant rule changes since the 1960s, when diminishing offense threatened the popularity of the game and resulted in the League shrinking the strike zone and lowering the pitching mound.  The goals of the most recent changes were to increase offense (again), speed up the game, and enhance player safety, especially during the Covid-19 pandemic.  The changes included a universal designated hitter, expanded postseason, increased base size, pitch clocks, banning of defensive shifts, and the introduction of the so-called ``ghost runner" rule for extra innings. Under the ghost runner rule, each team begins their half of extra innings with a runner (the ghost runner) on second base. Originally implemented in the Covid-shortened 2020 season and made permanent in 2023, the rule dramatically reduces the likelihood of long games.  In this paper, we show that teams are employing an inferior strategy with the ghost runner.  Specifically, we present clear evidence teams should sacrifice bunt more often in extra innings, even if they have to pinch hit for quality hitters who are not good bunters.  In general, sacrifice bunting is not an effective strategy \parencite{tango2007book} and its usage has decreased dramatically in the last two decades. However, there are specific cases when sacrifice bunting may be an effective strategy \parencite{evans_talk}.  This paper demonstrates another case when sacrifice bunting is the best strategy, suggesting potential overreach in the application of early sabermetrics results that have almost eliminated the strategy from the game. In addition, we use causal inference methodology to frame the analysis and estimate the effect of bunting in the home half of extra innings.  Therefore, a second contribution of this paper is to add to the growing body of literature using casual inference methodology to evaluate in-game strategies in professional sports.

A regulation MLB game consists of nine innings.  The away team bats first in the top half of each inning followed by the home team in the bottom half. If the game is tied at the end of regulation, innings continue until the winner is decided, either by the home team going ahead in an inning or the away team leading at the end of the inning. Since 2020, 10\% of games, or approximately 16 games per season for a typical team, went to extra innings. Despite playing a long season of 162 games, each game is very important to teams.  Assemble an average team expected to win half its games (i.e. 81 games) and find a way to eek out several more wins, your team is likely to make the playoffs where anything can happen in the best of five or seven game playoff series.  From 2021-2023, 93\% of teams with 90 or more wins (in other words, average teams that flipped at least nine of 81 losses to wins) made the playoffs.  While estimates vary, teams pay millions of dollars in player salary per statistical win above an average replacement \parencite{Cameron2009}. Winning just a couple more of these extra-inning games each season by changing strategies is worth millions of dollars in player salary.  

In this paper, we consider cases when the home team bats after the away team fails to score in the top half of an extra inning.  In the 2021-22 seasons, this occurred 249 times (about 4 times per team per season).  An obvious strategy is to sacrifice bunt on the first plate appearance. A bunt is a lightly tapped ball in front of home plate.  When executed properly, the defense does not have time to throw to third base to get the advancing ghost runner and must settle for getting the bunter out at first base.  With a runner on third base and one out, the home team is in a dominating position, with many ways to score the runner from third base to end the game.  However, home teams are choosing to swing away instead of bunting.  They bunted only 53 times (21.2\%) and swung away 196 times (78.8\%).  When they did bunt, the strategy was tremendously effective.  The home team won the game in the inning they bunted in 73.6\% of the cases.  When they swung away, they only won 56.6\% for an odds ratio for winning comparing bunting to swinging away of 2.13 (95\% C.I. 1.13, 4.30).   

By itself, the higher win percentage for bunting is not a strong argument to bunt more often. The observed association between bunting and winning is a biased estimate of the effect of bunting because the bunters are very different from the batters who swung away (who we refer to as `nonbunters').  The bunters have much more experience bunting likely making them more proficient at the task. We would not expect the nonbunters to have had the same success rate bunting as the bunters did.  The nonbunters are better hitters than the bunters making them more likely to end the game swinging away than the bunters would have. One approach to estimating an unbiased effect of bunting is to conduct a randomized controlled trial.  We could flip a coin before each plate appearance to determine whether the batter will be a bunter or nonbunter and then compare the results.  Unfortunately, we have yet to find a manager willing to enroll in our study.  

Instead, we turn to ideas in causal inference. The goal of causal inference is to estimate the effect of an intervention on one variable (the ``treatment") on another variable (the ``outcome").  In observational studies, measures of association relating the treatment to the outcome are often biased estimates of the effect when the treatment and outcome have common causes.  This bias is called confounding and the common causes are confounding variables.  Consider the case of two treatment groups. Confounding occurs when the process of assigning subjects to treatment groups is expected to result in groups that are unbalanced in ways that affect the outcome.  In our case, managers are more likely to select proficient bunters and weaker hitters to bunt, and vice versa with swinging away.  They may also consider other factors such as the opposing team's pitcher and defense.

The central idea of causal inference is that we can obtain unbiased estimates of effects (as if we had conducted a randomized experiment) from observational studies when a set of conditions hold.  These conditions are consistency, exchangeability, and positivity \parencite{hernan2010causal}.  Consistency means the treatment is a well-defined intervention that corresponds to values in the data. The exchangeability condition holds when the assignment of subjects to a treatment group depends only upon a set of measured covariates.  Positivity means there is a non-zero probability of receiving each treatment conditional upon the covariates.  In other words, everyone needs to have some chance of being a bunter and nonbunter based on the covariates measured. 

Causal inference has roots in economics, epidemiology, computer science, and biostatistics.  Joshua Angrist, Guido Imbens and David Card were awarded the Nobel Prize in economic sciences for their work in causal inference.  In sports analytics and statistics, there is growing interest in applying these methodologies to evaluating in-game strategies \parencite{szymanski2022anticipating}. \textcite{vock2018estimating} used the potential outcomes framework and $g$-computational formula to obtain estimates of a batter's performance if they had adopted the plate discipline of someone else.  Using data from the Japanese Baseball League, \textcite{nakahara2023pitching} used propensity score stratification to evaluate pitching strategies and found pitching outside the most effective.  In football, \textcite{yam2019lost} created a matched cohort of teams to evaluate fourth down decision-making and found evidence more aggressive teams did better. In soccer, \textcite{gauriot2019fooled} matched shots hitting the goal post to similar scoring shots and found managers and evaluators over-rewarded the lucky difference between the two.

The organization of this paper is as follows.  First, we define the decision to bunt or swing away in terms of the potential outcomes framework.  Next, we discuss the propensity score and the use of inverse probability weighting (IPW) in the outcome model. Then, we present results from the propensity score and  outcomes models.

\section{Data}


We obtained Retrosheet \parencite{retrosheet} play-by-play information for first plate appearances (PA) in the home half of tied extra-inning MLB games from the 2021-2022 seasons using the baseballr R package \parencite{petti_gilani_2021}.  We used the Chadwick tools \parencite{chadwick} to extract key fields from the event files. \textcite{albert2018analyzing} contains further background and R tools for analyzing Retrosheet data. The data includes a field $BUNT\_FL$ which indicates whether the plate appearance ended with a bunt in fair play. We retrieved yearly statistics for batters and pitchers from the Lahman database \parencite{lahman2022} and merged them to event records.

\section{Methods}

\subsection{Intervention and Outcome Variables}

The intervention in this study is bunting ($A = 1$) versus swinging away ($A = 0$) on the first plate appearance.  Labeling plate appearances as bunts or not is complicated by the fact that plate appearances usually consist of several pitches.  The batter may switch strategies during the plate appearance by attempting to bunt some pitches and swing away on others.  For example, a batter might attempt to bunt until they have two strikes, then swing away because a foul bunt is a strikeout.  Likewise, they might consider bunting but then change their mind after reaching a more favorable count for swinging away.  Furthermore, the field $BUNT\_FL$ only indicates whether the plate appearance ended in a bunt placed in fair territory, but does not distinguish cases when the batter attempted bunts on earlier pitches.  Together, these issues represent a potential violation of the consistency assumption of causal inference if we use $BUNT\_FL$ as the measure of the intervention. Specifically, if batters attempt to bunt some pitches and swing away on others, there are multiple versions of the intervention that do not correspond to values in the data (specifically, the $BUNT\_FL$ field). To investigate the scope of the violation, we randomly sampled 30 cases each of bunts and swinging away, and reviewed pitch-by-pitch behavior to assess how often batters switched strategies.  We found very few cases of strategy switching.  In extra innings, teams decide on a strategy for the plate appearance and stick with it.  Therefore, the violation of consistency is minor and we use the $BUNT\_FL$ field as our intervention measure with little impact of the validity of estimates.

The outcome of interest $Y$ is whether the home team wins the game in the inning of the intervention.  If the team does not win in the inning, the game continues to additional extra innings. Let $Y_1$ and $Y_0$ be the potential outcomes, or counterfactual outcomes, when the batter bunts and swings away, respectively.  Note we cannot observe both $Y_1$ or $Y_0$ for a plate appearance and cannot estimate $E(Y_1)$ and $E(Y_0)$ directly from the data.  The effect of interest $\theta_{BUNT}$ is the odds ratio for winning comparing bunting to swinging away.  
\[\theta_{BUNT} = \frac{E(Y_1) / (1-E(Y_1))}{E(Y_0) / (1-E(Y_0))}\] 
The unadjusted association $\theta_{Crude}$ below is a biased estimate of $\theta_{BUNT}$ due to confounding.  
\[\frac{E(Y|X=1)/(1-E(Y|X=1)}{E(Y|X=0)/(1-E(Y|X=0)}\]


\subsection{Propensity Score Model}
The propensity score is the probability of the batter bunting conditional on factors related to the manager's decision to bunt.  To estimate each batter's propensity score, we fit a logistic regression model predicting a bunt in the plate appearance using the following variables: OPS (on-base plus slugging), Sacrifice Bunt Rate per 100 PA, and the pitcher's ERA (earned run average). We chose these three statistical measures to capture overall hitter effectiveness, bunting frequency, and pitcher effectiveness. OPS measures both a player's ability to hit for power (slugging) and consistency (on-base percentage) \parencite{costa2019understanding}. We determined bunting aptitude through the rate of sacrifice bunts per 100 plate appearances. As for pitchers, ERA indicates their ability to limit runs against their opponents. Propensity scores close to one (zero) indicate the batter is likely (unlikely) to bunt.  


\subsection{Outcome Model}

To estimate the effect of bunting on the home team winning, we used the propensity scores in inverse probability weighting (IPW). The purpose of IPW is to create a pseudo population of plate appearances in which the bunters and nonbunters have similar distributions of important confounding variables (\textcite{austin2011introduction}).  The outcome model is a logistic regression model predicting winning conditional upon whether the batter bunted or swung away.  Each plate appearance is weighted by the inverse of the propensity score.  A common challenge with IPW is propensity scores close to zero result in very large weights when inverted.  We utilized the method of \textcite{crump2009dealing} and trimmed observations with propensity scores less than 0.1 and greater than 0.9.  To adjust for residual confounding, we also include OPS, sacrifice rate, and ERA in the IPW model.

\section{Results}

\subsection{Unadjusted Results}

Table 2 compares OPS, sacrifice rate, and pitcher ERA of the bunters and nonbunters. As expected, managers were more likely to bunt with batters who had more experience bunting and more likely to swing away with better hitters.  Interestingly, it does not appear that the quality of the pitcher impacted the decision the bunt.  Table 3  displays the distribution of results from plate appearances by whether the team bunted.  For example, one result is a runner on first and third with no outs. 
 Bunting was much more likely to have a favorable result.  We defined the following results as favorable: run scores, runners on first and third with no outs, runners on first and second with no outs, and a runner on third with one out.  All other results were considered unfavorable or were not observed in the data.


\subsection{Propensity Score Model}
\begin{figure}[htbp]  
\centering
\includegraphics[width=\textwidth]{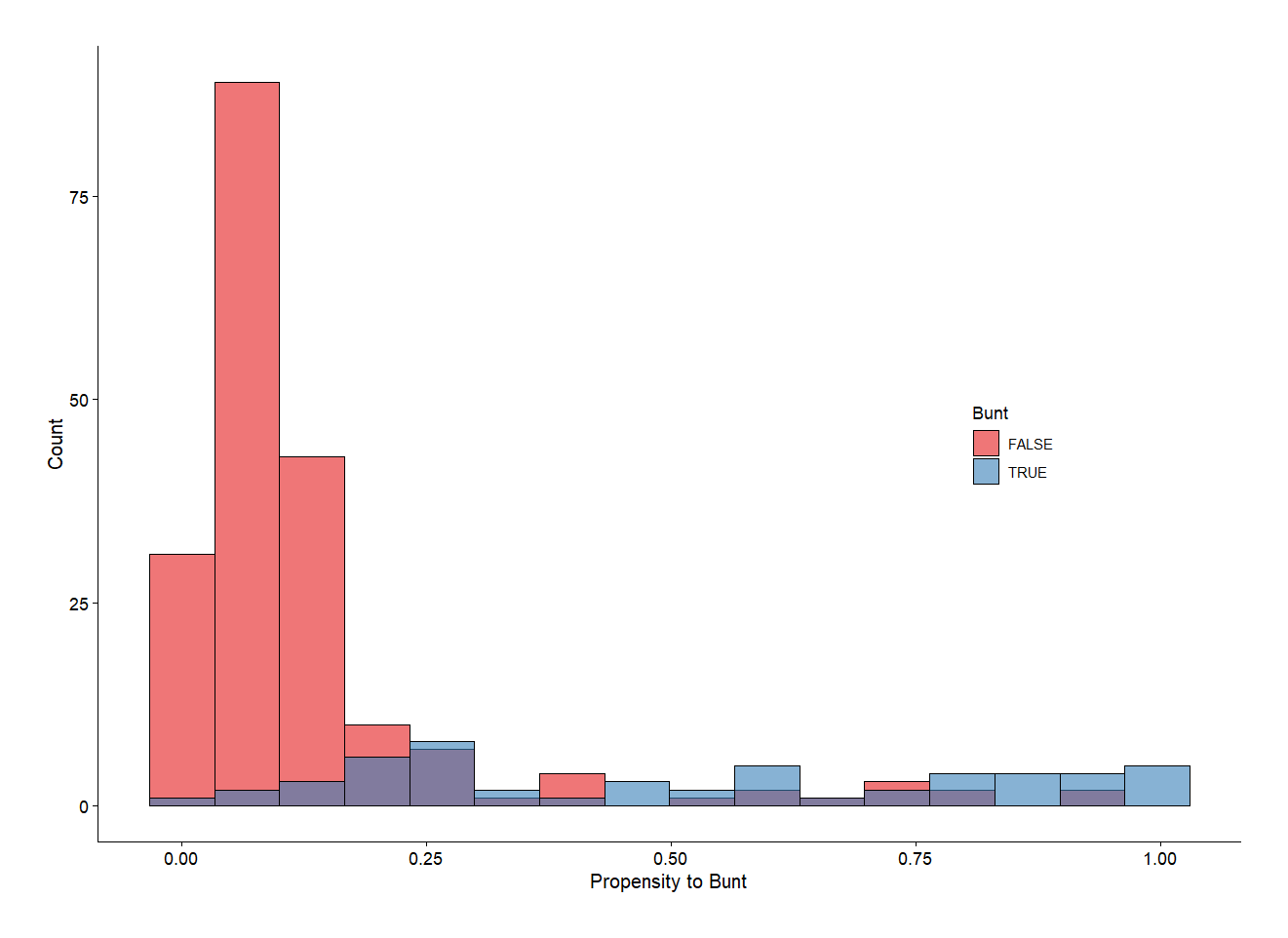}
\label{fig:example}
\end{figure}

{Figure \ref{fig:example}} depicts the distribution of propensity scores for bunters and nonbunters.  OPS and sacrifice rate were important predictors of bunting.  Plate appearance with propensity scores less than 0.1 or greater than 0.9 were trimmed from the data set and not included in the outcome model.  

\subsection{Outcome Model}
Table \ref{tab:outcome-model} odds ratios and confidence intervals from the  unadjusted and IPW models.  The IPW model had an odds ratio for winning comparing bunting to swinging away of 1.86 (95\% C.I.: 1.07, 3.27) after adjusting for hitting/bunting and pitcher quality, suggesting bunting is the more effective strategy.  This effect is smaller than the unadjusted odds ratio of 2.13. 

\section{Discussion}


This study joins a growing body of research refining the evidence of the effectiveness of playing so-called ``small ball".  Small ball refers to strategies and playing styles that emphasize speed, moving runners along, and using other creative means to generate runs beyond hitting for power and consistency. Early analyses in sabermetrics provided compelling evidence that sacrifice bunting and stealing bases, which are staples of small ball, were inferior strategies.  More specifically, they showed that sacrifice bunting and stealing bases, in general, decrease a team's run expectancy.  As a result, teams abandoned small ball strategies, and the sacrifice bunt and stolen base are now rare events in baseball.  However, these analyses looked at small ball strategies under general circumstances. ``[W]hen bunting, teams are often not as concerned about the overall run potential of an inning, but the probability of simply scoring one run'' \parencite{baseballanalysts}.  More recent studies have found the strategies to be effective in specific situations, suggesting there was an overreaction to the evidence presented. ``It turns out that while the initial observation that sacrifice bunting caused a decline in run expectation was not erroneous, it was foolish to extrapolate that the sacrifice bunt is always a bad play'' \parencite{baumer2014}. In our specific situation, with a runner on second base at the start of the inning and no outs, needing just one run to win the game, the concept of run expectancy loses some of its significance. This is because run expectancy typically assesses the potential for scoring multiple runs, whereas our objective is to score a single run. 

There are important limitations to our study.  For an unbiased bunting effect, we rely on the assumption of exchangeability.  Specifically, we assume that the decision to bunt only depends upon how frequently the batter bunted before (a proxy for how good of a bunter the batter is), how good the batter is when swinging away (as measured by OPS), and the quality of the pitcher (as measured by ERA).  Bunting skill is probably the most important factor in deciding whether to bunt.  However, there is not a good measure of bunting skill.  Future analyses could use location data to develop player-level bunting proficiency metrics to control for confounding in models assessing bunting strategies.  We could also look at other factors affecting the bunt decision such as pitcher-batter specific matchups. 



Our analysis suggests managers should pinch hit (i.e.\ substitute) good bunters for good hitters more often. One disadvantage of this strategy is the good hitter is no longer available for later in the game as he is not allowed to return to the game. However, it is rare to receive more than one plate appearance in extra innings. Over 90\% of extra inning games conclude within the first two extra innings. With the ghost runner, extra-inning games are typically decided in the 10th with less than 8\% going to 12 innings or more. 

This research offers invaluable insights for baseball managers, unveiling the strategic advantage of employing the bunting strategy in tied extra-inning scenarios when the home team is up to bat with the game in the balance. Managers can harness this newfound knowledge to make well-informed decisions that carry the potential to significantly impact the outcome of critical games. 




\section{Acknowledgements}

\section{References}

\normalem
\printbibliography

@article{austin2011introduction,
  title={An introduction to propensity score methods for reducing the effects of confounding in observational studies},
  author={Austin, Peter C},
  journal={Multivariate behavioral research},
  volume={46},
  number={3},
  pages={399--424},
  year={2011},
  publisher={Taylor \& Francis}
}

@article{yam2019lost,
  title={What was lost? A causal estimate of fourth down behavior in the National Football League},
  author={Yam, Derrick R and Lopez, Michael J},
  journal={Journal of Sports Analytics},
  volume={5},
  number={3},
  pages={153--167},
  year={2019},
  publisher={IOS Press}
}

@article{vock2018estimating,
  title={Estimating the effect of plate discipline using a causal inference framework: An application of the G-computation algorithm},
  author={Vock, David Michael and Vock, Laura Frances Boehm},
  journal={Journal of Quantitative Analysis in Sports},
  volume={14},
  number={2},
  pages={37--56},
  year={2018},
  publisher={De Gruyter}
}

@article{nakahara2023pitching,
  title={Pitching strategy evaluation via stratified analysis using propensity score},
  author={Nakahara, Hiroshi and Takeda, Kazuya and Fujii, Keisuke},
  journal={Journal of Quantitative Analysis in Sports},
  volume={19},
  number={2},
  pages={91--102},
  year={2023},
  publisher={De Gruyter}
}

@article{szymanski2022anticipating,
  title={Anticipating the honeymoon: Event study estimation of new stadium effects in Major League Baseball using the imputation method},
  author={Szymanski, Stefan},
  journal={Economic Inquiry},
  year={2022},
  publisher={Wiley Online Library}
}

@book{costa2019understanding,
  title={Understanding sabermetrics: An introduction to the science of baseball statistics},
  author={Costa, Gabriel B and Huber, Michael R and Saccoman, John T},
  year={2019},
  publisher={McFarland}
}

@misc{hernan2010causal,
  title={Causal inference},
  author={Hern{\'a}n, Miguel A and Robins, James M},
  year={2010},
  publisher={CRC Boca Raton, FL}
}

@misc{evans_talk,
    title = {Treatment Effect Heterogeneity in MLB Bunting Strategies.},
    author = {Evans, Katherine and Lopez, Michael},
    year = {2019},
    publisher = {Joint Statistical Meetings}
}

@article{gauriot2019fooled,
  title={Fooled by performance randomness: Overrewarding luck},
  author={Gauriot, Romain and Page, Lionel},
  journal={Review of Economics and Statistics},
  volume={101},
  number={4},
  pages={658--666},
  year={2019},
  publisher={MIT Press One Rogers Street, Cambridge, MA 02142-1209, USA journals-info~…}
}

@online{lahman2022,
    author = "Sean Lahman",
    title = "Sean Lahman's Baseball Database",
    year = "2022",
    url = "http://www.seanlahman.com/baseball-archive/statistics",
    note = "June 15, 2023"
}

@book{tango2007book,
  title={The book: Playing the percentages in baseball},
  author={Tango, Tom M and Lichtman, Mitchel G and Dolphin, Andrew E},
  year={2007},
  publisher={Potomac Books, Inc.}
}

@online{retrosheet,
  author    = {{Retrosheet}},
  title     = {{Retrosheet}},
  year      = {2023},
  url       = {https://www.retrosheet.org/},
  note      = {Accessed on June 15, 2023}
}

@online{chadwick,
  author    = {Ted Turocy},
  title     = {{Chadwick: Software Tools for Game-Level Baseball Data}},
  year      = {2023},
  url       = {https://chadwick.readthedocs.io/en/latest/},
  note      = {Accessed on June 15, 2023}
}

@book{albert2018analyzing,
  title={Analyzing baseball data with R},
  author={Albert, Jim and Baumer, Benjamin S},
  year={2018},
  publisher={Chapman and Hall/CRC}
}

@misc{petti_gilani_2021,
  author = {Bill Petti and Saiem Gilani},
  title = {baseballr: The SportsDataverse's R Package for Baseball Data.},
  url = {https://billpetti.github.io/baseballr/},
  year = {2021}
}

@online{baseballanalysts,
  author    = {Dan Levitt},
  title     = {{Empirical Analysis of Bunting}},
  year      = {2006},
  url       = {http://baseballanalysts.com/archives/2006/07/empirical_analy_1.php},
  note      = {Accessed on November 2, 2023}
}

@book{baumer2014,
  title={The Sabermetrics Revolution: Assessing the Growth of Analytics in Baseball},
  author={Baumer, Benjamin and Zimbalist, Andrew},
  year={2014},
  publisher={University of Pennsylvania Press}
}

@online{Cameron2009,
  author    = {Dave Cameron},
  title     = {Win Values Explained: Part Six},
  year      = {2009},
  url       = {https://blogs.fangraphs.com/win-values-explained-part-six/},
  journal   = {FanGraphs}
}

@article{crump2009dealing,
  title={Dealing with limited overlap in estimation of average treatment effects},
  author={Crump, Richard K and Hotz, V Joseph and Imbens, Guido W and Mitnik, Oscar A},
  journal={Biometrika},
  volume={96},
  number={1},
  pages={187--199},
  year={2009},
  publisher={Oxford University Press}
}

\section{Tables}

\begin{table}[htbp]
    \centering
    \label{tab:stats-comparison}
    \renewcommand{\arraystretch}{1} 
    \begin{tabular}{@{}lcc@{}}
        \toprule
        & \textbf{Bunt (X = 1), $n = 53$} & \textbf{Swings Away (X = 0), $n = 196$} \\
        \midrule
        \textbf{Home Team Wins (Y = 1)} & 73.6\% & 56.6\% \\
        \textbf{Batter OPS} & $0.629 \pm 0.09$ & $0.761 \pm 0.11$ \\
        \textbf{Batter Sacrifice Rate per 100 PA} & $0.86 \pm 0.80$ & $0.11 \pm 0.29$ \\
        \textbf{Pitcher ERA} & $3.80 \pm 2.3$ & $3.89 \pm 2.2$ \\
        \bottomrule
    \end{tabular}
    \caption{Comparison of Bunters and Nonbunters in the first plate appearance in the bottom half of extra innings.}
\end{table}

\begin{table}[htbp]
    \centering
    \label{tab:plate-results}
    \renewcommand{\arraystretch}{1} 
    \begin{tabular}{@{}lcccccc@{}}
        \toprule
        & \multicolumn{4}{c}{\textbf{Favorable result for Home Team}} & \multicolumn{1}{c}{\textbf{Unfavorable}} \\
        \cmidrule(lr){2-5} \cmidrule(lr){6-6}
        \textbf{} & \textbf{Runs Scored} & \textbf{1st/3rd No Outs} & \textbf{1st/2nd No Outs} & \textbf{3rd 1 Out} & \textbf{All Others} \\
        \midrule
        \textbf{Bunts (n=53)} & 0\% & 14.5\% & 0\% & 69.1\% & 16.4\% \\
        \textbf{Swings Away (n=196)} & 7.6\% & 9.6\% & 30.8\% & 21.2\% & 30.8\% \\
        \bottomrule
    \end{tabular}
    \caption{First plate appearance, regular season extra innings, 2021-2022}
\end{table}

\begin{table}[H]
    \centering
    \label{tab:outcome-model}
    \renewcommand{\arraystretch}{1} 
    \begin{tabular}{@{}lcc@{}}
        \toprule
        \textbf{Model} & \textbf{Odds Ratio} & \textbf{95\% Confidence Interval} \\
        \midrule
        Unadjusted & 2.13 & (1.13, 4.30) \\
        IPW & 1.86 & (1.07, 3.27) \\
        \bottomrule
    \end{tabular}
    \caption{Results of the Unadjusted and IPW Model}
\end{table}


\clearpage
\section{Figure Captions}
Figure 1: Bunt propensity distribution of the 249 batters in our data-set by bunting and swinging away
\clearpage
\section{Figures}

\end{document}